\documentclass[preprint,floatfix,superscriptaddress,showkeys,showpacs]{revtex4-1}
\usepackage{CJK}
\usepackage{graphicx}
\usepackage{dcolumn}
\usepackage{bm}
\usepackage{amsmath}
\usepackage{latexsym}
\usepackage{hyperref} 
\usepackage{booktabs}
\usepackage{array}

\usepackage{color,soul} 

\newcommand{\rfig}[1]{Fig.~\ref{#1}}
\newcommand{\rtab}[1]{Table.~\ref{#1}}
\newcommand{\rref}[1]{Ref.~\onlinecite{#1}}

\makeindex

\begin{document}

\title{Asymmetric Fraunhofer pattern in Josephson junctions from heterodimensional superlattice V$_5$S$_8$}
\begin{CJK*}{UTF8}{gbsn}
\author{Juewen Fan({\CJKfamily{gbsn}范珏雯})}
\author{Bingyan Jiang({\CJKfamily{gbsn}江丙炎})}
\author{Jiaji Zhao({\CJKfamily{gbsn}赵嘉佶})}
\author{Ran Bi({\CJKfamily{gbsn}毕然})}
\affiliation{State Key Laboratory for Artificial Microstructure and Mesoscopic Physics, Frontiers Science Center for Nano-optoelectronics, Peking University, Beijing 100871, China}
\author{Jiadong Zhou({\CJKfamily{gbsn}周家东})}
\affiliation{Key Laboratory of Advanced Optoelectronic Quantum Architecture and Measurement (Ministry of Education), Beijing Key Laboratory of Nanophotonics \& Ultrafine Optoelectronic Systems, and School of Physics, Beijing Institute of Technology, Beijing 100081, China}
\author{Zheng Liu({\CJKfamily{gbsn}刘政})}
\affiliation{School of Materials Science and Engineering, Nanyang Technological University, Singapore 639798, Singapore}
\author{Guang Yang({\CJKfamily{gbsn}杨光})}
\author{Jie Shen({\CJKfamily{gbsn}沈洁})}
\author{Fanming Qu({\CJKfamily{gbsn}屈凡明})}
\author{Li Lu({\CJKfamily{gbsn}吕力})}
\affiliation{Beijing National Laboratory for Condensed Matter Physics, Institute of Physics, Chinese Academy of Sciences, Beijing 100190, China}
\author{Ning Kang({\CJKfamily{gbsn}康宁})}
\affiliation{Key Laboratory for the Physics and Chemistry of Nanodevices and Department of Electronics, Peking University, Beijing 100871, China}
\author{Xiaosong Wu({\CJKfamily{gbsn}吴孝松})}
\email{xswu@pku.edu.cn}
\affiliation{State Key Laboratory for Artificial Microstructure and Mesoscopic Physics, Frontiers Science Center for Nano-optoelectronics, Peking University, Beijing 100871, China}
\affiliation{Collaborative Innovation Center of Quantum Matter, Beijing 100871, China}
\affiliation{Shenzhen Institute for Quantum Science and Engineering, Southern University of Science and Technology, Shenzhen 518055, China}

\begin{abstract}
Introduction of spin{\textendash}orbit coupling (SOC) in a Josephson junction (JJ) gives rise to unusual Josephson effects. We investigate JJs based on a newly discovered heterodimensional superlattice V$_5$S$_8$ with a special form of SOC. The unique homointerface of our JJs enables elimination of extrinsic effects due to interfaces and disorder. We observe asymmetric Fraunhofer patterns with respect to both the perpendicular magnetic field and the current. The asymmetry is influenced by an in-plane magnetic field. Analysis of the pattern points to a nontrivial spatial distribution of the Josephson current that is intrinsic to the SOC in V$_5$S$_8$.
\end{abstract}



\maketitle
\end{CJK*}

In the past decade, the interplay between spin{\textendash}orbit coupling (SOC) and superconductivity through the proximity effect has received significant attention\textsuperscript{\cite{Buzdin2008,Reynoso2008Sep,Lutchyn2010Aug,Bergeret2015,Wu2019}}. In materials with strong SOC, a variety of interesting effects may occur when combining them with superconductors, depending on the form of the SOC, structural symmetry, etc\textsuperscript{\cite{Buzdin2008,Liu2010,Yokoyama2014May,Yokoyama2014nov,Bergeret2015,Konschelle2015,Rasmussen2016}}. For instance, transition metal dichalcogenides with an Ising SOC can turn a conventional s-wave superconductor into an Ising superconductor through the proximity effect\textsuperscript{\cite{Wu2019}}. Another example, which has attracted enormous interest, is the construction of the Majorana zero mode in spin{\textendash}orbit coupled nanowires or topological insulators in close proximity to s-wave superconductors\textsuperscript{\cite{Lutchyn2010Aug,Oreg2010Oct,Fu2008,Lyu2018Oct,Banerjee2018Dec,Zheng2019Jun}}. In Josephson junctions (JJs), the combination of a Zeeman field and a Rashba spin{\textendash}orbit interaction can give rise to a supercurrent, even if the phase difference between two superconductors is zero, realizing a $\varphi_0$ junction\textsuperscript{\cite{Buzdin2008,Reynoso2008Sep,Bergeret2015}}.

A major playground for such studies is superconductor{\textendash}normal metal{\textendash}superconductor (SNS) JJs, in which the normal metal provides SOC. It is well-known that the critical current of a conventional JJ as a function of magnetic field exhibits a Fraunhofer interference pattern, described by\textsuperscript{\cite{Tinkham1996}} 
\begin{equation}
I_{\mathrm{c}}(B_z)=I_{\mathrm{c}}(0) \left| \frac{\sin(\pi B_z S/\Phi_0)}{\pi B_z S/\Phi_0}\right|
\label{eq.Fraunhofer}
\end{equation}
, where $I_\mathrm{c}$ is the critical current, $B_z$ the magnetic field perpendicular to the junction plane, $S$ the effective area of junction, and $\Phi_0=h/2e$ is the flux quantum. $I_\mathrm{c}$ is apparently independent of the polarity of $B_z$, $I_\mathrm{c}^\pm(B_z)=I_\mathrm{c}^\pm(-B_z)$, where $\pm$ indicates the current polarity. Its amplitude usually remains unchanged upon reversing the current $I$, too, $I_\mathrm{c}^+(B_z)=-I_\mathrm{c}^-(B_z)$. That is, the Fraunhofer pattern is symmetric in $B_z$ and $I$. However, recent studies have shown that these symmetries can be broken, especially in the presence of SOC. Broken current symmetry may be induced by structural asymmetry of the junction\textsuperscript{\cite{Hu2007Aug,Wu2021Mar,Misaki2021Jun}}, or a multiband effect\textsuperscript{\cite{Reynoso2008Sep,Yokoyama2014May,Trimble2021Jun}}. Broken field symmetry was attributed to disorder potentials\textsuperscript{\cite{Suominen2017}}, or nonuniform SOC\textsuperscript{\cite{Chen2018a,Assouline2019,Beach2021}}. Most of these cases require extrinsic mechanisms, which often exist in a practical junction. A general relation between the pattern asymmetry and the combination of a spin{\textendash}orbit coupling, in-plane magnetic field and disorder has been theoretically studied\textsuperscript{\cite{Liu2010,Rasmussen2016}}.

Here we study JJs fabricated with a novel material, $\mathrm{V}_5 \mathrm{S}_8$ of a heterodimensional superlattice, whose special form of SOC gives rise to a rare in-plane Hall effect\textsuperscript{\cite{Zhou2021}}. The homointerface of the junction enables investigation of intrinsic effects. It is found that both the magnetic field and current symmetries of the Fraunhofer pattern are broken. Moreover, the pattern can be tuned by an in-plane field. Consistent patterns and field dependence are observed in all devices. The phenomenon suggests a nontrivial spatial distribution of the Josephson current that is intrinsic to V$_5$S$_8$ and stems from its SOC.

Micro strips of V$_5$S$_8$ films were grown by chemical vapor deposition on SiO$_2$ substrates\textsuperscript{\cite{Zhou2021}}. Strips with about 10 nm thickness were selected for this study. As for superconductors, 31 nm aluminum, following a 2 nm titanium adhesion layer, were e-beam deposited. Some measurements were performed in an Oxford $^3$He cryostat with a base temperature of 250 mK. Experiments at a lower temperature were carried out in an Oxford dilution refrigerator with a base temperature of 10 mK. The magnetic field was provided by a three-axis vector superconducting magnet. To avoid the effect of hysteresis due to heating during a bias current sweep, the sweep always started from zero.
 
V$_5$S$_8$ film used in this study has a unique superlattice structure, in sharp contrast with conventional V$_5$S$_8$ compounds\textsuperscript{\cite{Zhou2021}}. It can be seen as VS$_2$ layers intercalated with V$_2$S$_2$ atomic chains. It exhibits an extraordinary in-plane Hall effect, suggesting a peculiar form of spin{\textendash}orbit-coupling. Because Al(31 nm)/V$_5$S$_8$(10 nm) bilayers are non-superconducting due to a strong inverse proximity effect, aluminum JJs can be made using the Al(31 nm)/V$_5$S$_8$(10 nm) bilayer as the normal metal weak link\textsuperscript{\cite{Fan2021}}. The JJ is a cross-bar structure made from a V$_5$S$_8$ strip and an aluminum strip. The resultant homointerface between the superconductor and normal metal is highly transparent and uniform, providing an ideal JJ platform for investigation of the intrinsic effect of SOC.

\begin{figure}[htbp]
	\begin{center}
		\includegraphics[width=1\columnwidth]{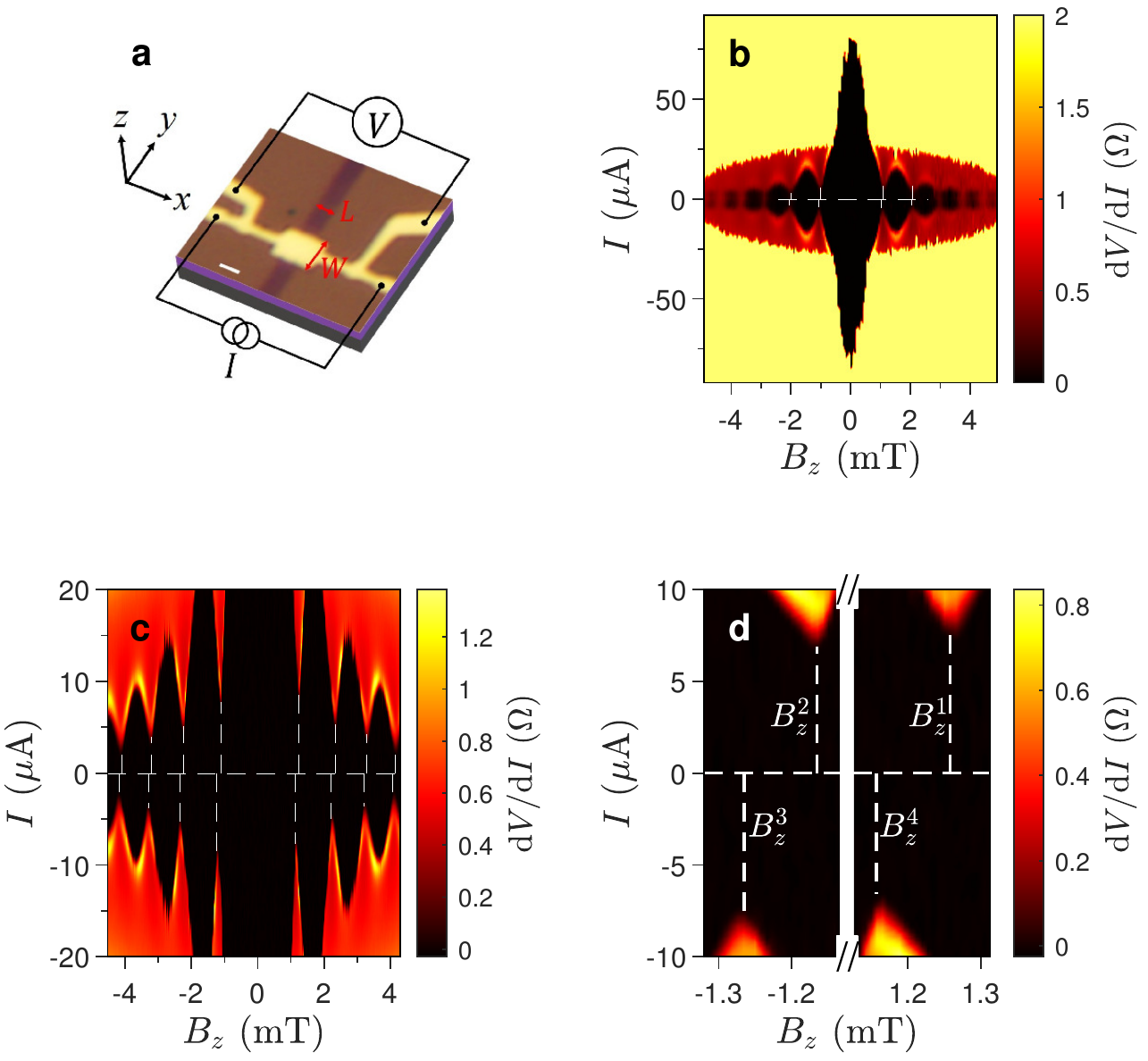}
		\caption{Fraunhofer interference patterns of a JJ. (a) An optical micrograph of the device with an illustration of the measurement configuration. The junction length is $L=0.92$ $\mu$m, the junction width is $W=1.75$ $\mu$m. The scale bar is 1 $\mu$m. (b) The interference pattern at 250 mK. The dashed white lines indicate the position of the nodal points and $I=0$. (c) The zoom-up pattern taken at 32 mK. (d) The further zoom-up scan on the first-order nodes in (c). A break in the $x$ axis is introduced for better comparison.}
		\label{fig.fraunhofer}
	\end{center}
\end{figure}

The measured JJ device is illustrated in \rfig{fig.fraunhofer}a. Since our superlattice V$_5$S$_8$ is a polar crystal, a coordinate system can be uniquely defined, which is necessary for the latter discussion. Note that the growth of V$_5$S$_8$ strips starts from a nucleation center and propagates along the $y$ axis. It is thus reasonable to assume that the growth is consistently towards the same crystalline direction. This direction is defined as the $-y$ direction. A positive current flows in the $+x$ direction of the crystal structure, while a positive magnetic field is in the $+z$ direction, which is pointing out of the plane of the substrate. The differential resistance $\mathrm{d}V/\mathrm{d}I$ displays a Fraunhofer-like pattern in the perpendicular magnetic field $B_z$ versus bias current $I$ mapping at 250 mK (\rfig{fig.fraunhofer}b). The period of the pattern $B_z^\mathrm{p}$ is 0.99 mT, yielding an effective junction area $S=\Phi_0/B_z^\mathrm{p}=2.09$ $\mu \mathrm{m}^2$, which is close to the nominal area of the JJ $S_\mathrm{n}=1.61$ $\mu \mathrm{m}^2$. The discrepancy comes from the London penetration depth and flux focusing. The diffusion constant of junction, $D$, is 6.9$\times 10^{-3}$ m$^2 \cdot$s$^{-1}$, calculated by $D=\frac{1}{3}(\frac{\pi k_\mathrm{B}}{e})^2 \frac{\sigma}{\gamma}$\textsuperscript{\cite{Pippard1960}}. Here, $\sigma=3.9 \times 10^7$ S$\cdot$m$^{-1}$ is the electrical conductivity of the weak link, $\gamma=1.4 \times 10^{2}$ J$\cdot$m$^{-3}\cdot$K$^{-2}$ is the electronic specific heat coefficient of Al\textsuperscript{\cite{Kittel2005}}. Since the electrical conductivity of Al film is much larger than that of V$_5$S$_8$, the electronic specific heat coefficient of the weak link can be approximate to that of Al. Then the mean free path $l_\mathrm{e}$ is 10.2 nm through $D=v_\mathrm{F} l_\mathrm{e}/3$, where $v_\mathrm{F}=2.02\times 10^8$ cm$\cdot$s$^{-1}$ is the Fermi velocity of Al\textsuperscript{\cite{Kittel2005}}. Because $l_\mathrm{e}$ is much shorter than the junction length $L=0.92$ $\mu$m, the junction is in the diffusive limit, in which case the Thouless energy $E_\mathrm{Th}$ is given by $\hbar D/L^2$\textsuperscript{\cite{Dubos2001}} and becomes 5.4 $\mu$eV, much smaller than the superconducting gap, indicating that the JJ is in the long junction limit.

The critical current at the low-order nodal points does not vanish as anticipated by a standard Fraunhofer pattern, which will be discussed later. The most unusual feature is that the pattern is lack of the mirror symmetries with respect to both the current and the field, $I_\mathrm{c}^{+}(B_z) \neq -I_\mathrm{c}^{-}(B_z)$ and $I_\mathrm{c}^{\pm}(B_z) \neq I_\mathrm{c}^{\pm}(-B_z)$. The nodal points in the upper half-plane ($I^+$) are not aligned with those in the lower half-plane ($I^-$). The misalignment can be seen more clearly in a zoom up mapping taken at low temperatures, where the superconducting transition is sharp. As depicted in \rfig{fig.fraunhofer}c, the nodal points in the upper half-plane shift to the right with respect to those in the lower half-plane. A further zoom up scan on the first-order nodes shows that aside from the horizontal shift, the nodal points also shift vertically (\rfig{fig.fraunhofer}d). These in the second and forth quadrants, assigned numbers of $n=2$ and 4, respectively, are closer to the $I=0$ axis than those in the first and third quadrants, assigned numbers of $n=1$ and 3. In all of 7 devices we have measured, the nodal points shift towards the same direction, suggesting that the effect is intrinsic to V$_5$S$_8$, rather than due to the sample geometry, differences at the interface or disorder. On the other hand, the inversion symmetry $-I_\mathrm{c}^{+}(B_z) = I_\mathrm{c}^{-}(-B_z)$ is retained.

\begin{figure}[htbp]
	\begin{center}
		\includegraphics[width=1\columnwidth]{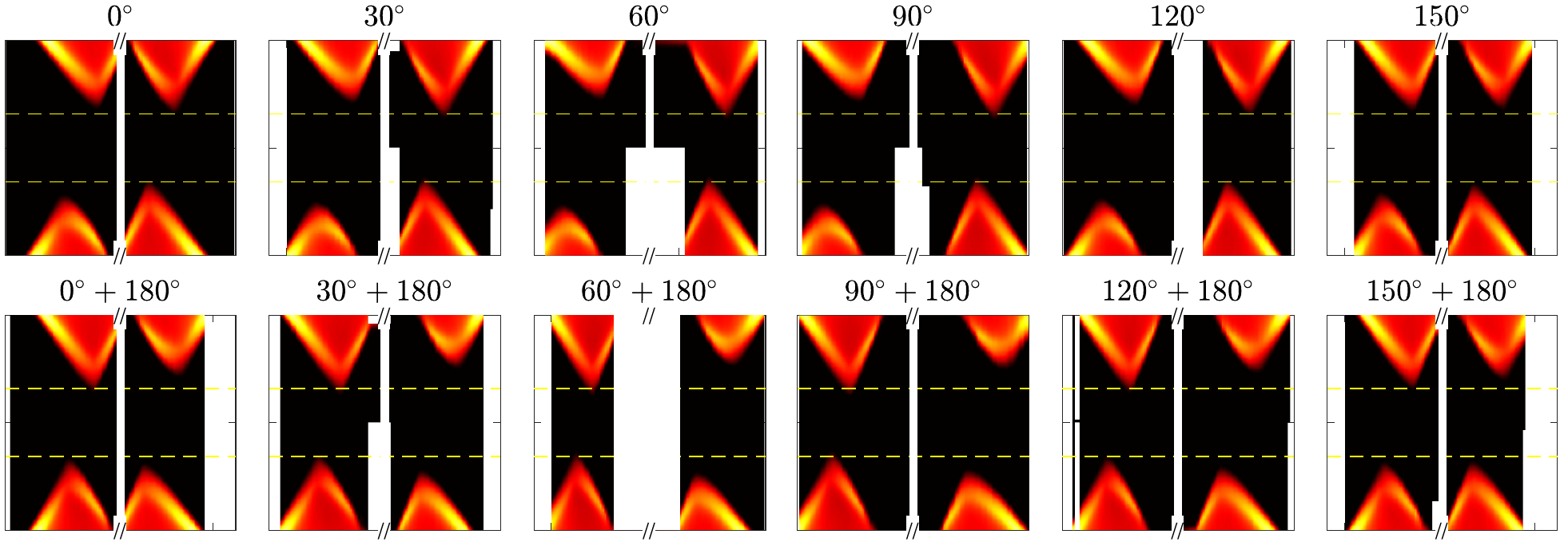}
		\caption{Fine mapping of the first-order nodes in the interference pattern in the presence of $B_\parallel=0.12$ T oriented at different angles with respect to the $x$ axis. Two dashed yellow lines at $I=\pm 3.158$ $\mu$A are used as a baseline to highlight the shift of the critical current at the nodal points.}
		\label{fig.finemaps}
	\end{center}
\end{figure}

\begin{figure}[htbp]
	\begin{center}
		\includegraphics[width=0.5\columnwidth]{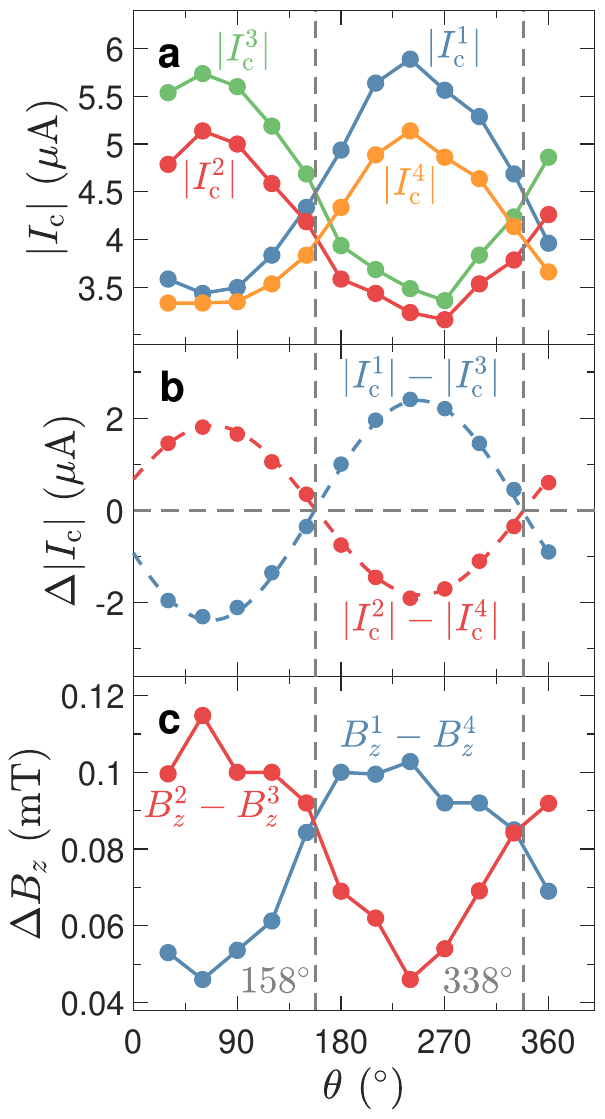}
		\caption{Angular dependence of the position of the nodal points. (a) Absolute value of the critical current. (b) The difference of the critical currents, $|I_\mathrm{c}^1|-|I_\mathrm{c}^3|$ and $|I_\mathrm{c}^2|-|I_\mathrm{c}^4|$. The dashed curves represent fits to a sine function. (c) The difference of the field at the nodal points, $B_z^2-B_z^3$ and $B_z^1-B_z^4$.}
		\label{fig.angulardep}
	\end{center}
\end{figure}

We now apply an in-plane field and investigate the change in the nodal point. \rfig{fig.finemaps} shows 12 fine mappings around the nodal points in \rfig{fig.fraunhofer}d in the presence of the in-plane field of $B_{\parallel}=0.12$ T along different directions. Let $\theta$ denote the angle between $B_{\parallel}$ and the $x$ axis. One may immediately notice that the inversion symmetry of pattern is now broken. The critical currents at four nodal points, $I_\mathrm{c}^n$ $(n=1,2,3,4)$, are plotted as a function of the field angle $\theta$, shown in \rfig{fig.angulardep}a. $I_\mathrm{c}^n$s oscillate with a $2\pi$ period. $|I_\mathrm{c}^1(\theta)|$ and $|I_\mathrm{c}^3(\theta)|$ are exactly out-of-phase, so are $|I_\mathrm{c}^2(\theta)|$ and $|I_\mathrm{c}^4(\theta)|$. Only at two critical angles, $\theta=158^{\circ}$ and $338^{\circ}$, $|I_\mathrm{c}^1|$ is equal to $|I_\mathrm{c}^3|$ and $|I_\mathrm{c}^2|$ is equal to $|I_\mathrm{c}^4|$, indicating restoration of the inversion symmetry. To quantify the degree of inversion symmetry breaking, let us employ the difference of $I_\mathrm{c}$ between a pair of nodes that are inversion symmetric in the absence of an in-plane field. As illustrated in \rfig{fig.angulardep}b, the asymmetry is at its maximum at $\theta=68^{\circ}$ and $248^{\circ}$. Similarly, we can assess the asymmetry from the field of nodal points. Let $B_z^n$ $(n=1,2,3,4)$ denote the perpendicular field at the nodal point. To eliminate the estimation error of the zero field due to unavoidable sample misalignment, we calculate $\Delta B_z^{23}=B_z^2-B_z^3$ and $\Delta B_z^{14}=B_z^1-B_z^4$, plotted as a function of $\theta$ in \rfig{fig.angulardep}c. The more $\Delta B_z^{23}$ deviates from $\Delta B_z^{14}$, the more asymmetric the pattern is. The evolution of the asymmetry is consistent with that inferred from $I_\mathrm{c}$. Comprehensive measurements of the angular dependence of the pattern on another sample, shown in the supplementary material, yield an almost identical behavior and a similar critical angle at which $|I_\mathrm{c}^1|$($|I_\mathrm{c}^2|$) intercepts $|I_\mathrm{c}^3|$($|I_\mathrm{c}^4|$). 

Since $|I_\mathrm{c}^1(\theta)|$ and $|I_\mathrm{c}^3(\theta)|$ oscillate with the same amplitude and opposite phase, i.e., $|I_\mathrm{c}^1(\theta)| = |I_\mathrm{c}^3(\theta+\pi)|$, we have $I_\mathrm{c}^1(B_{\parallel}) = -I_\mathrm{c}^3(-B_{\parallel})$, which, combined with the similar behavior of $\Delta B_z^{23}$ and $\Delta B_z^{14}$, suggests
\begin{equation}
I_\mathrm{c}^{+}(B_z,B_{\parallel})=-I_\mathrm{c}^{-}(-B_z,-B_{\parallel})
\label{eq.inversion}
\end{equation}
. In another word, the diffraction pattern is inversion symmetric if the total magnetic field is considered. This can also be directly seen in \rfig{fig.finemaps}, where the pattern at $\theta$ is inversion symmetric to that at $\theta+180^{\circ}$.

First, let us discuss the reason for the symmetry breaking at zero in-plane field seen in \rfig{fig.fraunhofer}. One possible cause is the self-field effect\textsuperscript{\cite{Barone1982}}. For JJs with certain geometries, a nonuniform current distribution in the superconductor electrodes can produce an additional field through the junction area, which shifts the pattern horizontally. As the self-field depends on the polarity of the current, the patterns in the upper and lower half-plane shift towards opposite directions. The geometry of our JJs has a mirror symmetry about the center line along the current direction, precluding the self-field effect. Even if the device slightly deviates from the mirror symmetry due to imperfections introduced by fabrication, the skewness of the pattern should vary from device to device. The consistent shift in all devices indicates that the symmetry breaking is associated with the crystal structure of V$_5$S$_8$.

Abrikosov vortices in superconducting electrodes near the JJ can lead to an asymmetric Fraunhofer pattern, too. However, the effect breaks the field symmetry, $I_\mathrm{c}^{\pm}(B_z) =I_\mathrm{c}^{\pm}(-B_z)$, leaving the current symmetry, $I_\mathrm{c}^{+}(B_z) =-I_\mathrm{c}^{-}(B_z)$, intact\textsuperscript{\cite{Golod2010,Golod2019}}, which is at odds with our results. Moreover, we intentionally made the superconductor leads narrow enough so that vortices can not enter\textsuperscript{\cite{Stan2004}}. For comparison, when the leads are wider than 0.65 $\mu$m, jittering of the pattern due to vortex jump started to appear in the Fraunhofer pattern.

Rasmussen \textit{et al.} have studied the interference pattern of a SNS JJ theoretically and related its symmetry to certain symmetries in the Hamiltonian\textsuperscript{\cite{Rasmussen2016}}. The symmetries that are required to break in order to break the mirror symmetries, $I_\mathrm{c}^{\pm}(B_z) =I_\mathrm{c}^{\pm}(-B_z)$ and $I_\mathrm{c}^{+}(B_z) =-I_\mathrm{c}^{-}(B_z)$, are listed in the \rtab{t2}. Here $\sigma_{x,y,z}$ are the Pauli matrices, $P_{x(y)}$ is the $x(y)$-parity operation, $T$ is the time reversal operation. Disordered potential $V_{x(y)}$, which is asymmetric under $P_{x(y)}$, unlikely plays an important role, otherwise the pattern in different devices would shift in random directions\textsuperscript{\cite{Suominen2017}}, not to mention the fact that our JJs, made from a continuous aluminum film, are presumably quite uniform. In the absence of an in-plane field, two forms of SOC, Rashba SOC $\alpha$ and Dresselhaus SOC $\beta$, can break most of the symmetries. It is plausible that V$_5$S$_8$ has both $\alpha$ and $\beta$ or similar forms of SOC, as those SOCs are the key ingredients in the theories on the in-plane Hall effect\textsuperscript{\cite{Zhang2011,Ren2016a,Liu2018a,Zyuzin2020}}, which has been observed in V$_5$S$_8$. However, there are still two more symmetry operators left, $\sigma_z P_y P_x$ and $\sigma_z P_x P_y T$, which guarantee the mirror symmetries of the pattern.

\begin{table}[htbp]
\centering
        \caption{Symmetry operations $U$ protecting $-I_\mathrm{c}^{+}(B_z)=I_\mathrm{c}^{-}(B_z)$ and $I_\mathrm{c}^{\pm}(B_z)=I_\mathrm{c}^{\pm}(-B_z)$ respectively adapted with permission from \rref{Rasmussen2016}.Copyrighted by the American Physical Society.}
        \begin{center}
        \begin{tabular}{p{2cm}p{4cm}p{2cm}p{4cm}}
            \hline
            \multicolumn{2}{l}{$-I_\mathrm{c}^{+}(B_z)=I_\mathrm{c}^{-}(B_z)$} & \multicolumn{2}{l}{$I_\mathrm{c}^{\pm}(B_z)=I_\mathrm{c}^{\pm}(-B_z)$}\\
            \cline{1-2}
            \cline{3-4}
            $U$& Broken by&$U$& Broken by\\
            \hline
            $P_y P_x$& $\alpha$, $\beta$, $V_x$, $V_y$&$\sigma_x P_y$ & $B_y$, $\alpha$, $V_y$\\
            $\sigma_z P_y P_x$  & $B_x$, $B_y$, $V_x$, $V_y$ &$\sigma_y P_y$  & $B_x$, $\beta$, $V_y$\\
            $\sigma_x P_y T$ & $B_x$, $\alpha$, $V_y$ &$P_x P_y T$  & $B_x$, $B_y$, $\alpha$, $\beta$, $V_x$, $V_y$\\
            $\sigma_y P_y T$ & $B_y$, $\beta$, $V_y$ &$\sigma_z P_x P_y T$ & $V_x$, $V_y$\\
            \hline
        \end{tabular}  
        \end{center}
        \label{t2}
\end{table}

It has been proposed that the mirror symmetries of the interference pattern, $I_\mathrm{c}^{\pm}(B_z) =I_\mathrm{c}^{\pm}(-B_z)$ and $I_\mathrm{c}^{+}(B_z) =-I_\mathrm{c}^{-}(B_z)$, are broken in SNS based on inversion symmetry breaking topological materials\textsuperscript{\cite{Chen2018}}. This is because the topological surface states on two sides of the JJ have different Fermi velocities, resulting in different Josephson currents. The interference pattern by the surface states is thus shifted and skewed, depending on the polarity of the current. Taken into account the dominant contribution from the bulk, the resultant inversion symmetric pattern agrees with what we observed. Broken mirror symmetries of the Fraunhofer pattern were observed in topological materials\textsuperscript{\cite{Bocquillon2016,Kononov2020Jun}}. The difference of the Josephson currents on two sides breaks the mirror symmetry of the Josephson current spatial distribution about the center line of JJ along the current. It is worth noting that this mirror symmetry is essential for the nodal point being on the $I=0$ axis. Therefore, the finite critical current of the nodal point in our data provides additional evidence, independent of the model on inversion symmetry breaking topological materials, that the Josephson currents on the two sides are not equal.

The asymmetry of the pattern is tuned by the in-plane field, seen in \rfig{fig.finemaps}. It has been shown that an in-plane field can suppress the Josephson current in the center of a JJ via a peculiar flux focusing effect\textsuperscript{\cite{Suominen2017}}. Consequently, the zero field critical current is reduced and the interference pattern becomes more like that of a superconducting quantum interference device. It works when the field is parallel to the current and does not depend on the field polarity. As the field is rotating in the plane, a period of $\pi$ is expected in the angular dependence of the critical current, in contrast to a period of $2\pi$ shown in \rfig{fig.angulardep}. Furthermore, the effect does not depend on the polarity of the current, so it will not break aforementioned two mirror symmetries of the pattern. At last, the flux focusing effect maximizes at $\theta=0^\circ$ and $180^\circ$, whereas the critical currents of the nodal points at these two angles are at neither the maximum nor the minimum, shown in \rfig{fig.angulardep}. These distinctions rule out the flux focusing effect of the in-plane field.

An in-plane field usually acts on electrons by the Zeeman effect, thus its strong influence on the pattern suggests that the asymmetry is related to the spin. In the case of inversion symmetry broken topological materials, the Zeeman effect will break the time reversal symmetry of the surface states, which alters the Josephson current on the two sides. The change in the current distribution inevitably leads to a change of the pattern. The critical angles of $\theta=158^{\circ}$ and $338^{\circ}$, at which the inversion symmetry of the pattern is restored, are probably related to the particular form of SOC in V$_5$S$_8$.

The unusual symmetry in the Fraunhofer pattern of Al{\textendash}(Al/V$_5$S$_8$){\textendash}Al Josephson junctions suggests an anomalous Josephson coupling stemming from the SOC in heterodimensional superlattice V$_5$S$_8$. This newly discovered compound, with its exotic SOC and the ability of forming a homointerface Josephson junction, offers an intriguing platform for investigation of unconventional Josephson effects and superconductivity.

\section*{Supporting Information}
The measurements of the in-plane field dependence of the Fraunhofer pattern on another Al{\textendash}(Al/V$_5$S$_8$){\textendash}Al Josephson junction S2.

\begin{acknowledgements}
Project supported by the National Key Basic Research Program of China (Grant No. 2016YFA0300600) and the National Natural Science Foundation of China (Grant Nos. 11574005 and 11774009).
\end{acknowledgements}

\providecommand{\newblock}{}

\pagebreak
\widetext
\begin{center}
\textbf{\large Supplemental Materials}
\end{center}
\setcounter{equation}{0}
\setcounter{figure}{0}
\setcounter{table}{0}
\setcounter{page}{1}
\makeatletter
\renewcommand{\theequation}{S\arabic{equation}}
\renewcommand{\thefigure}{S\arabic{figure}}
\renewcommand{\bibnumfmt}[1]{[S#1]}
\renewcommand{\citenumfont}[1]{S#1}

This Supplemental Material Section contains the measurements of the in-plane field dependence of the Fraunhofer pattern on another Al{\textendash}(Al/V$_5$S$_8$){\textendash}Al Josephson junction S2.

\begin{figure}[htbp]
	\begin{center}
		\includegraphics[width=1\columnwidth]{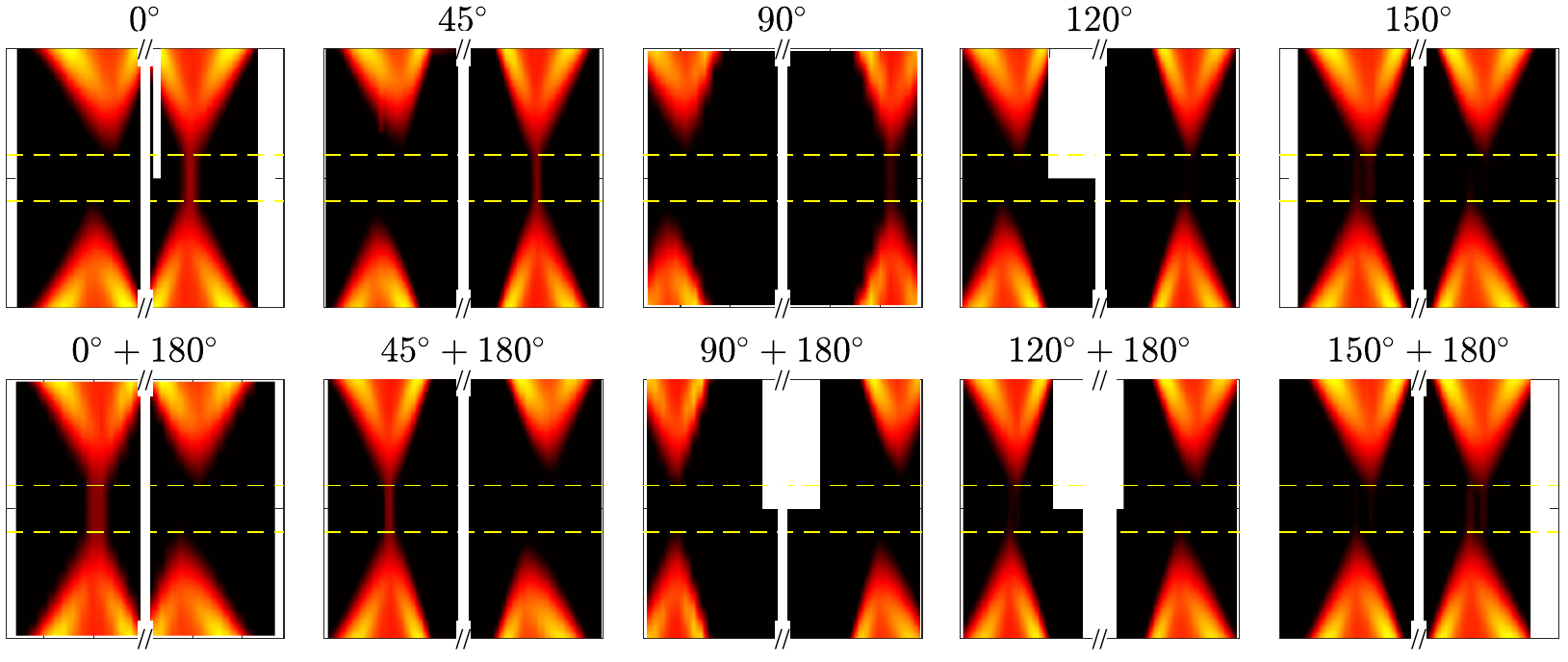}
		\caption{Fine mapping of the first-order nodes in the interference pattern of JJ S2 in the presence of $B_\parallel=0.12$ T oriented at different angles with respect to the $x$ axis. Two dashed yellow lines at $I=\pm 0.9$ $\mu$A are used as a baseline to highlight the shift of the critical current at the nodal points.}
		\label{fig.finemaps.s2}
	\end{center}
\end{figure}

\begin{figure}[htbp]
	\begin{center}
		\includegraphics[width=0.5\columnwidth]{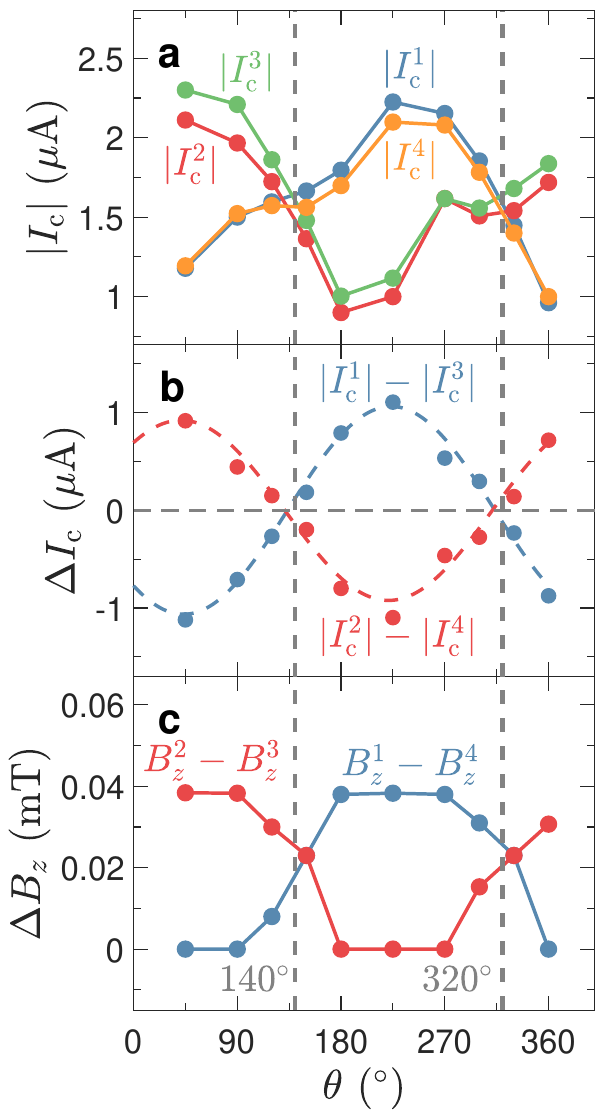}
		\caption{Angular dependence of the position of the nodal points of JJ S2. (a) Absolute value of the critical current. (b) The difference of the critical currents, $|I_\mathrm{c}^1|-|I_\mathrm{c}^3|$ and $|I_\mathrm{c}^2|-|I_\mathrm{c}^4|$. The dashed curves represent fits to a sine function. (c) The difference of the field at the nodal points, $B_z^2-B_z^3$ and $B_z^1-B_z^4$.}
		\label{fig.angulardep.s2}
	\end{center}
\end{figure}

JJ S2 in \rfig{fig.angulardep.s2} displays a behavior similar to that of the JJ in the main text, including the critical angles at which $|I_\mathrm{c}^1|$($|I_\mathrm{c}^2|$) intercepts $|I_\mathrm{c}^3|$($|I_\mathrm{c}^4|$). For JJ S2, the critical angles are about $140^{\circ}$ and $320^{\circ}$.

\clearpage

\end{document}